\documentclass[10pt]{article}
\usepackage{geometry}                
\usepackage{graphicx}
\usepackage{amssymb}
\usepackage{epstopdf}
\usepackage{comment,color}
\usepackage{hyperref}
\usepackage[utf8]{inputenc}
\usepackage{verbatim}
\title{Extragalactic plus Galactic model for IceCube neutrino events}
\author{Andrea Palladino\footnote{Gran Sasso Science Institute, andrea.palladino@gssi.infn.it}, Francesco Vissani\footnote{LNGS and Gran Sasso Science Institute, francesco.vissani@lngs.infn.it}}

\date{}                                           
\graphicspath{{figures/}}
\begin{document}
\maketitle
\begin{abstract}
The hypothesis that high energy cosmic neutrinos are power law distributed 
is critically analyzed. 
We propose a model with two-components that explains better the observations. The extragalactic component of the high energy neutrino flux 
has a canonical $E_\nu^{-2}$ spectrum while the   galactic component has a $E_\nu^{-2.7}$ spectrum;  both of them are significant. This model has several implications, that can be tested by IceCube and ANTARES in the next years. Moreover, the existence of a diffuse component, close to the Galactic plane
and that yields  (20-30)\% of IceCube's events, is interesting for the future $\mbox{km}^3$ neutrino telescopes located in the Northern Hemisphere and for gamma ray telescopes aiming at measuring events  till few 100 TeV from the Southern sky.
\end{abstract}
\section{Introduction}
Before IceCube, the most popular expectation on ultra high energy neutrinos, typically adopted in sensitivity studies,  was that 
1) most of them have an extragalactic origin, as  
motivated by the existence of many extreme astrophysical objects and of extragalactic cosmic rays; 
2) their spectrum is distributed like $E_\nu^{-2}$, as motivated by the expectations on cosmic rays at the sources and by
Fermi's acceleration mechanism. 
A particular very well-known implementation was the one due to Waxman \& Bahcall \cite{Waxman:1998yy}.

After IceCube, this hypothesis has been minimally modified in order to take into account that the {\em observed} spectrum 
does not agree well with $E_\nu^{-2}$. 
\textcolor{black}{The minimal modification amounts to use 
single power law $E^{-\alpha}$,  
isotropic and with $\alpha\sim 2.5$, a value obtained by fitting the 
observations of IceCube.}
In this work, pluses and the minuses of this position are summarized, motivating an alternative and more satisfactory hypothesis to account for the observations of IceCube and to forecast future findings.  If the alternative hypothesis is correct,  it is premature to dismiss the position described just above. Moreover, as we will discuss below, there are a lot of  interesting consequences for the astronomy of high energy neutrinos.

{\color{black}Before proceeding, two important clarifications are in order,\\
1) The aim of the present paper is not to prove wrong the assumption of a single power law $E^{-\alpha}$, isotropically distributed, that is most commonly adopted for the analyses of IceCube data. This  is still viable at present, if one adopts very conservative criteria (i.e., one discards some data set or information and/or if one accepts various discrepancies). 
Our aims are to illustrate the reasons of interest toward another hypothesis capable to interpret the IceCube data, illustrating its advantages and its physical interest,  deriving its implications, showing that it can be tested with future data.\\
2) Also the issue of minimality requires additional discussion. 
The ``single power law" hypothesis has two free parameters,
the normalization and the slope, that adds only one parameter to the 
previously adopted hypothesis (the normalization of the assumed   $E_\nu^{-2}$ flux).  
However, in view of the fact that, to date,  
we lack firm theoretical bases for the interpretation of IceCube observations, we should not exaggerate the heuristic power of  mathematical criteria. Moreover, in the simplest version of our model, that we adopt, the spectrum has two components with fixed slopes: one of them behaves as $E^{-2}_\nu$, the other one as $E^{-2.7}_\nu$. 
Thus the free parameters for the fit are again two, the normalizations of the fluxes.}

The structure of this paper is as follows. In Sect.~\ref{cd} we 
examine the hypothesis of single component for the high energy neutrinos, stressing its shortcoming. In Sect.~\ref{nh} we motivate and present the new hypothesis that includes two-components: one due to extragalactic neutrinos the other one due to galactic neutrinos. 
We quantify there the relative intensity and discuss the spectra. In 
Sect.~\ref{pre} we illustrate the predictions of our model for existing and future neutrino telescopes and the implications for gamma astronomy at the highest energy. A brief summary is contained in Sect.~\ref{summary}.

\section{Critical discussion of the single power law model\label{cd}}
In this section we proceed to a critical assessment of the single power law model for high energy neutrino emission, that, in view of its popularity and of the almost general acceptance at present,  can be considered as the {\em null hypothesis} adopted for data analysis.
\subsection{Angular anisotropy \label{boranga}} 
The list of the candidate extragalactic sources of ultra-high-energy neutrinos comprises several disparate astrophysical objects, including various classes of active galactic nuclei (AGN)  \cite{agn1,agn2} such as BL Lacertae objects \cite{bll1,bll2,bll3}; of peculiar galaxies, such as starburst-galaxies \cite{sb1,sb2,sb3}; of extreme stellar objects as Gamma ray bursters (GRB) \cite{grb1,grb2}; 
etc. They differ greatly in physical and observable characteristics. 
However, a  common feature of these objects is to have an almost isotropic angular distribution in the sky, making a possible exception for the 
brightest among them, which could stand out.   
This is the main reason why it is assumed that, in first approximation, the (supposedly extra-galactic) 
high energy neutrinos seen by IceCube should be isotropically distributed (but note incidentally that also certain hypothetical sources  in the Galaxy such as the halo might lead to 
a similar distribution). 

In reference \cite{Neronov:2015osa} the angular distribution of the events observed by IceCube was analyzed, showing a significant excess of events on the galactic plane. The analysis is repeated with different angular bins and the significance is never less than 3$\sigma$.
A similar work is done in the paper \cite{Troitsky:2015cnk}, where the arrival directions of high-energy neutrinos are discussed, focusing on the possibility that neutrinos are not only extragalactic but they can be also produced in the halo or in the disk of our Galaxy. The result of ref.~\cite{Troitsky:2015cnk} is that the present data are compatible with a purely extragalactic component, but also a mixed flux (extragalactic plus galactic neutrinos) is well compatible with the data. On the contrary a purely galactic component is disfavored at about 2$\sigma$, as  can be  seen in the Fig.~2 of that paper. 
{\color{black} Later in this paper, Sect.~\ref{angola}, 
we perform a new independent analysis of the angular distribution,
obtaining results consistent with those described here.}

The angular anisotropy represents the first hint to reconsider the null hypothesis. 


\subsection{Spectral distribution}
\label{falfa}
{\color{black}
In this section of the text, we discuss the fact that different datasets collected by IceCube suggest different
power law distributions. We begin with an introduction to 
the relevant datasets, where we indicate how they correspond 
to the neutrino fluxes, see Sec.~\ref{ce1}. 
Then, we will compare the power law distributions,
obtained from the different datasets, in Sec.~\ref{ce2}.

\subsubsection{The different datasets and their summaries as power-law distributions \label{ce1}}

IceCube has observed two main classes of events. 
The high energy starting events (HESE), whose vertex is contained in the detector, and the passing muons (aka throughgoing muons, aka tracks) that are the traditional signal of neutrinos. The HESE are divided in two classes: 1) 	\textit{direction}: those from the North sky and those from the South sky 2) \textit{topology}: those of shower type and those of track type. The passing muons instead are all from the North sky and all of track type. Note that the experimental sample of HESE North events is not very large, and the subset of HESE North tracks is still smaller;
the experimental sample of passing muon events, collected in IceCube, is much larger than the sample of HESE North muons. 
(We discuss below and in details which are the specific sets that we use in the analysis.)

The above considerations concern the experimental classification of the data. Passing now to the 
physical interpretation of these data, in terms of 
neutrino fluxes, it is essential to remark that, 
\begin{quote}
the passing muon events, measured in IceCube, give us information on the flux of muon neutrinos and antineutrinos
coming from the North sky, just as the set of HESE North muon events. 
\end{quote}
Let us recall the well-known fact that neutrino oscillations
connect tightly see e.g.~\cite{natural} the electron, the  muon and the tau neutrino fluxes,
and in the particular and natural case when the neutrinos derive from pion decays, these fluxes are approximately the same.
Let us finally recall that the inclusion of passing muon data has been important to verify the consistency of  
IceCube observations with three flavor neutrino oscilllations in the recent past 
\cite{Palladino:2015zua}.
All these considerations point out the importance of including the passing muons in the analysis.

\subsubsection{Comparison of the power law distributions obtained from different datasets \label{ce2}}

Let us begin discussing the HESE events.}
The all-flavor extraterrestrial component of the HESE observed by IceCube, after 4 years of data taking, 
is fitted by a single power law spectrum, without an energy cutoff, obtaining:
\begin{equation}
\frac{d \phi}{d E_\nu}=F \cdot 10^{-18} \frac{1}{\rm GeV \ cm^2 \ sec \ sr} \ \bigg(\frac{E_\nu}{100 \rm \ TeV} \bigg)^{-\alpha} 
\label{unit}
\end{equation}
with $F=6.7^{+1.1}_{-1.2}$ and $\alpha=2.50 \pm 0.09$ \cite{Aartsen:2015knd}.
However, an attentive analysis of the most recent IceCube data reveal a difference between the flux that comes from the Southern sky and the one that comes from the Northern sky. Indeed these two subsets of data give  
the following best-fit fluxes: $F_{\mbox{\tiny N}}=2.1^{+2.9}_{-1.6}, \ \alpha_{\mbox{\tiny N}}=2.0^{+0.3}_{-0.4}, \   
F_{\mbox{\tiny S}}=6.8^{+1.6}_{-1.5}, \ \alpha_{\mbox{\tiny S}}=2.56^{+0.11}_{-0.12} $.
\textcolor{black}{In order to quantify the difference between North and South HESE power-law fluxes we analyze the distribution of their  spectral indices, described approximately by two Gaussian functions $g_{\mbox{\tiny N}}(\alpha)$ and $g_{\mbox{\tiny S}}(\alpha)$.}
Following the procedure reported in the Appendix a mild tension of 1.3$\sigma$ between the dataset of HESE North and HESE South, $\delta \alpha = 0.56 \pm 0.42$ is found.

%

\textcolor{black}{Let us discuss now the passing muons. 
Since the IceCube experiment is located in the South pole,  the passing muons  give information on the muons neutrino flux coming from the North sky. This information must be consistent with the information from HESE North dataset, that are also generated by neutrinos that come from the Northern sky. Moreover, as already recalled, the passing muons dataset contains a larger number of events  than the HESE North dataset alone, which allows us to obtain stronger inferences. More precisely, 
comparing the flux obtained fitting the passing muons with  the flux  obtained fitting the HESE seen from the Southern sky results in} a much greater tension. 
In \cite{Aartsen:2015zva} the flux of diffuse astrophysical muon is analyzed and it is found $\alpha_p=1.91 \pm 0.20$ for the spectral index. \textcolor{black}{This flux is obtained using events 
in the energy range between 170 TeV and 3.8 PeV, but if the single power law hypothesis is the true one this flux \textit{must be valid} also at lower energies.} On the other side, the dataset of HESE South gives a preference for a softer spectrum, i.e. $\alpha_s \simeq 2.56 \pm 0.12$. 

\textcolor{black}{
Using the  procedure described in the Appendix, it is possible to compare the fluxes that describe the 
various datasets, just as we did above with the fluxes obtained only from HESE data. The results are,}
\begin{itemize}
\item A discrepancy 
equal to 2.8$\sigma$ when we 
compare the fluxes obtained from the passing muons with the one obtained from HESE-Southern sky, namely, $\delta \alpha =0.65 \pm 0.23$.
\item A similar tension of 2.7$\sigma$ between the power-law distribution that describes the passing muons dataset and the 
global best fit of HESE, namely, $\delta \alpha = 0.59 \pm 0.22$.
\item Compatibility between HESE North and passing muons, namely, $\delta \alpha =0.09 \pm 0.45$.
\end{itemize}
In a recent talk, presented by Schoenen (IceCube Collaboration) at ``TeV Particle Astrophysics 2015, Kashiwa, Japan" the 29th of October, a similar result is obtained: the disagreement between the tracks detected in 6 years and the combined analysis of 4 years is claimed to have the  significance of 3.6$\sigma$ that is a bit stronger and thus even more significant than the results we have just illustrated. 

Since the tracks are generated by muon neutrinos\footnote{The contribution of the tracks that come from $\nu_\tau$ is of the order of 5$\%$, because we have to take into account not only the branching ratio but also the energy of the secondary muon with respect to the energy of the primary neutrino.} and the showers by all flavor neutrinos, the discrepancy may be explained in 
two major ways,
\begin{enumerate}
\item The flux of $\nu_\mu$ has a different behavior with respect to the flux of $\nu_e$ and $\nu_\tau$, but there are no theoretical reasons to sustain this point; 
\item The Northern sky and the Southern sky are detecting two different population of neutrinos: an almost purely extragalactic component from the Northern sky and a mixed component (galactic plus extragalactic) from the Southern sky. 
\end{enumerate}

In conclusion, analyzing the spectral distribution of the fluxes derived by the different datasets it is possible to conclude that there is a difference in the shape between the neutrino flux observed in the Southern sky and the one observed in the Northern sky, with a significance of at least 2.7$\sigma$.  \textcolor{black}{Combining the 1.3$\sigma$ (HESE North - HESE South) and 2.7$\sigma$ (Passing muons - HESE South) hints, the overall significance becomes 3.0$\sigma$, which means a p-value less than 0.3\%}.

  This significance 
represents the second hint to go beyond the single power law model. 

\subsection{Extragalactic power law neutrinos, protons and $\gamma$-rays}
In this section we analyze the connections between cosmic neutrino and other radiations, i.e.\ cosmic rays and gamma.
\begin{figure}[t]
\centering
\includegraphics[scale=0.7]{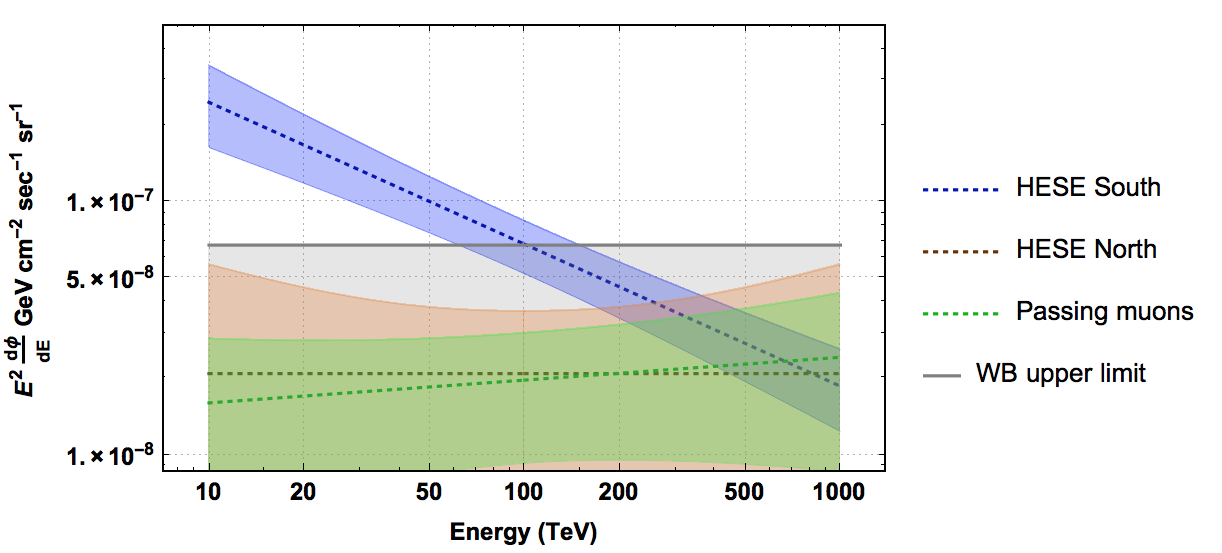}
\caption{\textit{Comparison between the limit given by a source transparent to ultra-high energy cosmic rays and the power law fluxes that describe the present IceCube data.}}
\label{WBlimit}
\end{figure}
Let us consider a source transparent to protons, as it happens in the model proposed by Waxman and Bahcall \cite{Waxman:1998yy}. This model predicts an upper bound for the cosmic neutrino flux, connected to the amount of energy of the cosmic rays, in the energy interval between $10^{19}-10^{21}$ eV. Their argument goes as follows: High energy protons, produced by the extra galactic sources, lose a fraction $\epsilon$ of their energy through photo-meson production of pions before escaping the source, the resulting present day energy density of all-flavor neutrinos will be given by,
\begin{equation}
E_{\nu}^2 \frac{d\phi}{dE_\nu} \simeq \frac{3}{2} \ \epsilon \ I_{\rm max}
\end{equation}
where $I_{\rm max}$ is given by:
\begin{equation}I_{\rm max}= 0.25 \xi_{Z} t_H \frac{c}{4 \pi} E_{CR}^2 \frac{d \dot{N}_{CR}}{d E_{CR}} \simeq 1.5 \times 10^{-8} \xi_Z \ \rm GeV cm^{-2} s^{-1} sr^{-1} 
\end{equation}
where $\xi_Z$ is connected with cosmology and takes into account the energy loss by neutrinos at redshift different from 0 and the upper bound is obtained for $\epsilon=1$, i.e. when all the energy of protons is transferred to pions. The numerical value of $I_{\rm max}$ is obtained using $E_{CR}^2 \frac{d \dot{N}_{CR}}{d E_{CR}} = 10^{44} \rm erg Mpc^{-3} yr^{-1}$. Comparing this model with the fluxes that have been derived by IceCube using the 4 years dataset, the results illustrated in the Fig.~\ref{WBlimit} are obtained. \textcolor{black}{The bands related to the fluxes are due to the uncertainties on the normalization and on the shape. The upper limit is given by,
\begin{equation}
\phi+\Delta \phi= \phi + \sqrt{[\phi(F,\alpha)-\phi(F^+,\alpha)]^2+[\phi(F,\alpha)-\phi(F,\alpha^+)]^2}
\end{equation}
where $\phi$ is the flux at best fit, \textcolor{black}{$F^+=F+\Delta F$} and $\alpha^+=\alpha+\Delta \alpha$ if $E_\nu < \rm 100 \ TeV$ or $\alpha^+=\alpha-\Delta \alpha$ if $E_\nu > \rm 100 \ TeV$. With the same procedure, the lower limit $\phi-\Delta \phi$ is obtained. \textcolor{black}{The values of $F$, $\alpha$ and their uncertainties are reported in section \ref{falfa}.}
} \\
\textcolor{black}{The two fluxes derived from HESE-Northern sky and from the passing muons dataset are compatible with the \textcolor{black}{Waxman-Bahcall} model.
On the other hand, the data from HESE-Southern sky begin to exhibit some critical behavior below 60-70 TeV,
because they exceed this limit. Note that in this region, a lot of events were already observed, so it cannot be a problem attributed to the low statistic. \textcolor{black}{To have a more precise idea, let us use the flux that was derived by the IceCube collaboration in \cite{Aartsen:2015knd}, including events above 10 TeV. The energy flux of HESE from Southern sky at 25 TeV is equal to:
\begin{equation}
E_\nu^2 \frac{d\phi_{\mbox{\tiny S}}}{dE_\nu}(25 \rm \ TeV)=1.5 \pm 0.4 \times 10^{-7} \rm \ \frac{GeV}{cm^2 \ s \ sr}
\end{equation}
It means that the Waxman and Bahcall limit is violated at $2\sigma$ at that energy and the violation is even larger at lower energies: in fact, the flux should extend also at lower energies, if the hypothesis of single power law holds true.} Once again, also these considerations suggest to reconsider some of the assumed hypotheses. }

Another connected remark emerges comparing the flux of neutrinos with the IGRB (intergalactic gamma-ray background).
The first step in this direction was done by Murase et al. \cite{Murase:2013rfa}\cite{Murase:2014tsa}, using the two years dataset of IceCube events. In this work the $pp$ scenario is considered and the neutrino flux is compared to the diffuse $\gamma$-ray background flux measured by Fermi. A strong upper limit on the spectral index of the source of neutrinos it is obtained, i.e. $\alpha \leq 2.2$. A softer spectrum would give an excess of neutrinos with respect to the $\gamma$-ray in the 100 GeV region and this is in tension with the theory. 
Also in the paper of Bechtol et al.\cite{Bechtol:2015uqb} this kind of analysis is performed using more recent data. The extrapolated flux of neutrinos exceeds the flux of gamma around 0.1 TeV, even in the most conservative hypothesis in which the extrapolation follows an $E^{-2}$ spectrum. This excess represents a problem in the scenario of star-forming galaxy, where the $pp$ interaction is the dominant mechanism for the production of neutrinos. Moreover in that energy region the gamma are not absorbed, even if they come from distance of some Gigaparsec. 

\textcolor{black}{The above remarks 
are our last hint that the null hypothesis requires}  some critical consideration. 
The previous issues, concerning the comparison of  
power law distributed neutrinos  with 
protons and $\gamma$-rays, can indicate some of the following
possibilities, \\
\textit{i)} an exotic source of neutrinos (e.g.\ exotic dark matter);  \\
\textit{ii)} absorption of protons and/or gammas into the source\footnote{It could happen especially in the $p\gamma$ mechanism of production, because the $\gamma$-rays produced by the decay of $\pi^0$ can interact itself with the target photons contained into the source and they can be trapped into the source. In this kind of opaque source the connection with neutrinos can be difficult, as discussed in  \cite{Murase:2015xka}.}; \\
\textit{iii)} a galactic component of neutrinos on top of the extragalactic component. \\
In this paper we will explore the last option. It should be stressed that the limit on the neutrino flux, given by the comparison with cosmic rays and IGRB, only applies on the extragalactic neutrinos.

\section{The hypothesis of two-components spectrum
\label{nh}}
Considering the hints of the previous sections, we explore the  hypothesis that the Northern sky is seeing an almost purely extragalactic flux of neutrinos and the Southern sky is seeing both extragalactic and galactic neutrinos. This hypothesis can justify the spectral differences in the IceCube dataset \textit{and} the angular anisotropy of the observed events. Moreover this hypothesis ``saves" the upper limit of Waxman and Bahcall, that is in strong disagreement with the IceCube global fit already at $\simeq 100$ TeV. At the same time, it solves the problems arising from the comparison with the IGRB. 

\subsection{General motivations}

Our hypothesis is motivated by the position of the Earth in the Galaxy and by the location of IceCube detector. Let us begin with a standard model for the Galaxy, given by a cylinder of radius 15 kpc and a variable height of some kpc. 
The axis of rotation of the Earth forms an angle of 30$^{\rm o}$ with the galactic plane, so the equator of the Earth makes an angle of 60$^{\rm o}$ with it. The equator is represented by a plane, perpendicular to the normalized vector $(u_x,u_y,u_z)=(0.484,0.747,0.456)$. 
Imposing the condition that the plane includes the Earth, that is at 8.5 kpc from the center of the Milky Way,
the equation of the plane is obtained. 
\begin{figure}[t]
\centering
\includegraphics[scale=0.4]{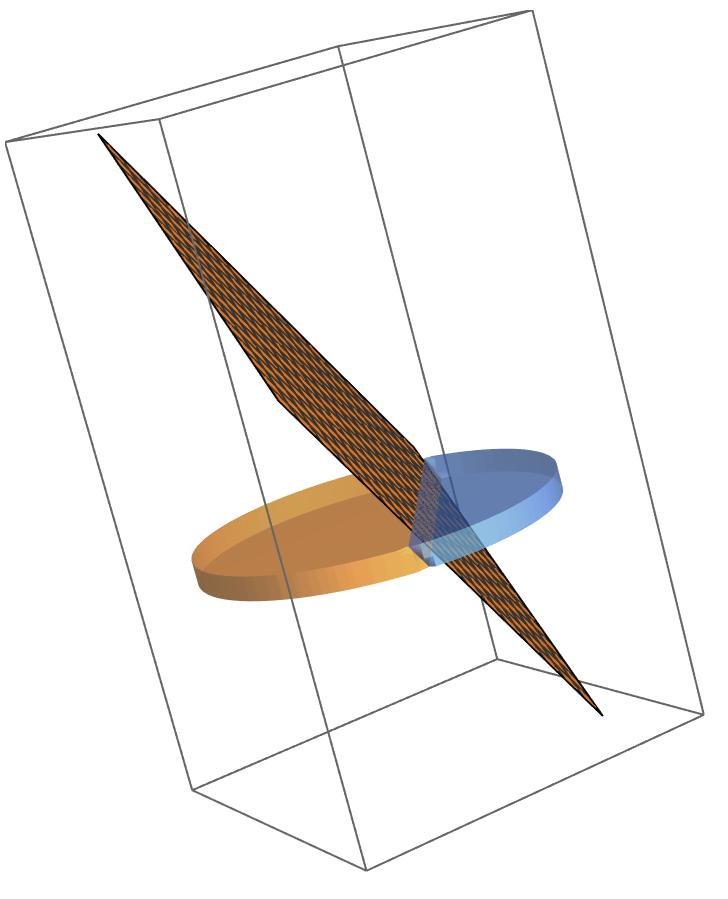}
\caption{\textit{A simplified cylindrical model of the 
Galaxy. The relative orientation of the  Earth's equatorial plane (orthogonal to rotation axis) is also shown.}}
\label{gal}
\end{figure}
At this point it is easy to  evaluate the volume of the Galaxy seen by the Northern sky and the one seen by the Southern sky; the result is about 70$\%$ from the South and $30\%$ from the North. 

In addition to this, one can take into account that the matter is not distributed uniformly, as can be seen, e.g., from Fig.~2 of  
\cite{Vissani:2011vg}. This is another good reason to believe that the Southern sky is seeing also a galactic component of neutrinos, that is on the contrary negligible from the Northern sky. 

However, the above modeling is good only for a cursory exploration
and to obtain a first idea. The region where the Galactic neutrino emission is more intense is not known a priori and should be eventually measured by neutrino observatories. E.g., it is natural to modify and extend the above model of the region where neutrino emission is more intense,  by adding two parameters: 
the radius of the cylinder and its galactic height.\footnote{An alternative parameterizations uses two exponential distributions
$n_{\nu}\propto \exp(-r/\delta r) \exp(-|h|/\delta h)$ where $r$ is the distance from the Galactic center and $h$ is the Galactic height;
$\delta r$ and $\delta h$ are model parameters.} Of course if the 
radius is well below  the distance from the center, $R\sim 8.5$ kpc, 
all the emitting region will lie in the Southern  hemisphere. The values of these parameters have to be obtained from the analysis of
 high energy neutrino data.

Note incidentally that the Fermi satellite sees an intense, diffuse galactic emission till 100 GeV. 
At higher energies, closer to those explored by neutrino telescopes, 
HESS has proved that the galactic neutrino emission due to point sources is quite intense; a diffuse galactic component at high energy, unfortunately, is not yet probed and hard to be probed. 
\textcolor{black}{In short, the existing gamma ray data indicate the existence of a significant component of the high energy radiation that can be attributed to the Galaxy, and this consideration does not contradict (but rather supports) the hypothesis that something similar happens for high energy neutrinos.}
We will come back on these considerations later on.


\subsection{Galactic component}

\begin{figure}[t]
\centering
\includegraphics[scale=0.65]{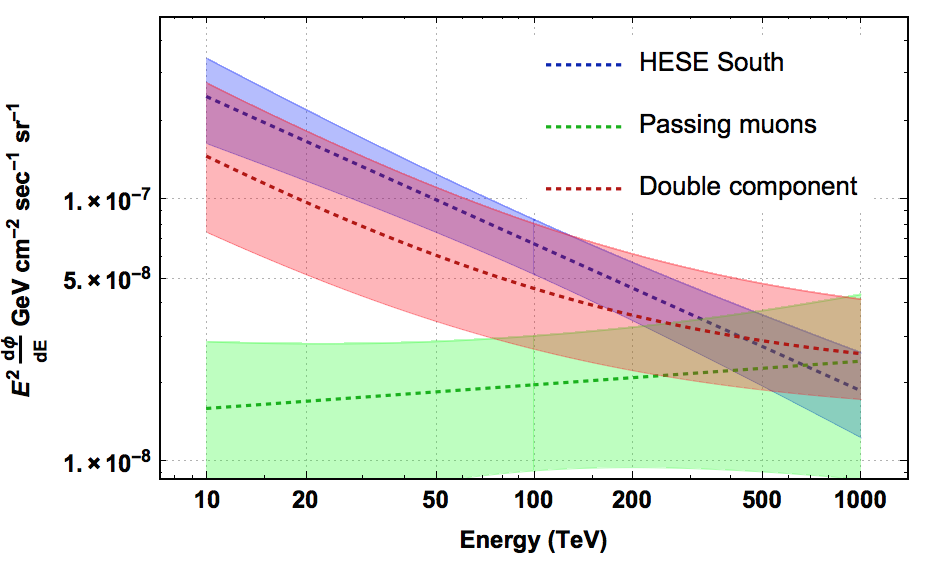}
\caption{\textit{Comparison between the fluxes measured by IceCube from the Northern sky and Southern sky with the two-components flux.}}
\label{eeg}
\end{figure}

\subsubsection{Spectrum and intensity of the galactic component}
Elaborating the hypothesis that IceCube is observing two populations of neutrinos, it seems natural to  describe the neutrinos that come from the Northern sky, that are assumed to be extragalactic,
 with an unbroken power law, distributed isotropically over the sky. 
 Moreover, it becomes natural to describe the observations of IceCube by adding a new a power law flux in the Southern sky, in which the galactic population of neutrinos gives an important contribution.
For the extragalactic component, the spectral index $\alpha=2$ is adopted, since it coincides with the value measured for the Northern sky flux at the best fit. For the galactic component, the spectral index $\alpha=2.7$ is adopted, i.e., the value of the galactic cosmic rays. It follows that:
\begin{eqnarray}
\frac{d \phi_{\mbox{\tiny N}}}{d E_\nu}&=&\frac{d \phi_{\mbox{\tiny EG}}}{d E_\nu} \nonumber \\
\frac{d \phi_{\mbox{\tiny S}}}{d E_\nu}&=& \frac{d \phi_{\mbox{\tiny G}}}{d E_\nu}+\frac{d \phi_{\mbox{\tiny EG}}}{d E_\nu} \nonumber
\end{eqnarray}
or more explicitly:
\begin{eqnarray}
&\displaystyle \frac{d \phi_{\mbox{\tiny N}}}{d E_\nu}=&F_{\mbox{\tiny EG}}  \ \bigg(\frac{E_\nu}{100 \rm \ TeV} \bigg)^{-2} \cdot 10^{-18} \frac{1}{\rm GeV \ cm^2 \ sec \ sr} \\
&\displaystyle \frac{d \phi_{\mbox{\tiny S}}}{d E_\nu}=&\bigg[F_{\mbox{\tiny G}}  \ \bigg(\frac{E_\nu}{100 \rm \ TeV} \bigg)^{-2.7} +F_{\mbox{\tiny EG}}  \ \bigg(\frac{E_\nu}{100 \rm \ TeV} \bigg)^{-2} \bigg]  \cdot 10^{-18} \frac{1}{\rm GeV \ cm^2 \ sec \ sr}
\label{2compeq}
\end{eqnarray}
Considering the events with deposited energy above 60 TeV,\footnote{Following \cite{Palladino:2015zua} we limit the analysis to this subset of data, where the atmospheric background gives a small contribution.} IceCube has seen 32 events in 4 years, 9 from the Northern sky, 22 from the Southern sky and one at latitude zero. Using the effective areas for each hemispheres, that recently became available on the IceCube website \cite{ice3}, the likelihoods for the extragalactic and the galactic normalizations ($F_{\mbox{\tiny EG}}, \ F_{\mbox{\tiny G}}$) can be obtained. The likelihood for the normalization at 100 TeV for the flux from the Northern sky, is defined as:
\begin{equation}
\mathcal{L}_{\mbox{\tiny EG}}(s) \propto (b+3.6 s)^{9.5} \cdot \exp[-(b+3.6 s)]
\end{equation}
where $b=2.1$ is the background due to conventional atmospheric neutrinos, considering the best-fit value\footnote{We scaled in time the expectation for the backgrounds contained in \cite{Aartsen:2014gkd}. We are not considering the charm component, that is zero at the best fit.}. The coefficient 3.6, instead, is the number of events expected in the case of normalization equal to one at 100 TeV. The result is $F_{\mbox{\tiny EG}}=2.1^{+1.1}_{-0.6}$ at 68$\%$ of confidence level. This extragalactic component is perfectly compatible with the upper limit given by the IGRB  observed by Fermi (see \cite{Bechtol:2015uqb}). \textcolor{black}{Let us remark that we are using the HESE both for the North and South hemisphere, because for this dataset the effective areas are available and we can perform our calculation using $\alpha=2$ for the extragalactic component. The inclusion of the passing muons could certainly reduce the uncertainties on the galactic component that we will calculate in a while, but it is not possible at the moment; for this reason, our results are conservative.}

\begin{table}[t]
\caption{\textit{Comparison of the two component model 
with the events from the Southern sky with deposited energy above 60 TeV of the 4 years dataset of IceCube.  For the galactic and the extragalactic components the best-fit value are used, see Eq.~\ref{chefig}. For the prompt neutrinos the number is at 90$\%$ C.L. and the best-fit value is 0.}}
\begin{center}
{
\begin{tabular}{|c|c|c|c|c|c|c|}
\hline
& \textbf{\footnotesize Gal. $E^{-2.7}$} & \textbf{\footnotesize Extragal. $E^{-2}$} & \textbf{\footnotesize Backg.} & \textbf{\footnotesize Prompt} & \textbf{\footnotesize Expected} & \textbf{\footnotesize Observed} \\
\hline
\textbf{\footnotesize Tracks} & 1.2 & 2.5 & 1.5 & $<$0.8 & 5.2  & 4.5  \\
\hline
\textbf{\footnotesize Showers} & 6.5& 10.1 &0.5 & $<$3.1 & 17.2 & 18 \\ 
\hline
\textbf{\footnotesize Total} & 7.7 &12.6 & 2.0 & $<$3.9  & 22.3 & 22.5 \\
\hline 
\textbf{\footnotesize Fraction} & 34.5$\%$ & 56.5$\%$ & 9$\%$ & & & \\
\hline
\end{tabular}}
\end{center}
\label{default}
\end{table}

Taking into account the uncertainties given by the supposed extragalactic component, the likelihood for the normalization at 100 TeV of the galactic component, seen from the Southern sky, is given by:
\begin{equation}
\mathcal{L}_{\mbox{\tiny G}}(s) \propto \int_0^{\infty} (b+6.0 s_{eg} + 3.1 s)^{22.5} \cdot \exp[-(b+6.0s_{eg}+3.1 s)] \cdot \mathcal{L}_{\mbox{\tiny EG}}(s_{eg}) d s_{eg} \label{stimagal}
\end{equation}
where the background is equal to the previous case and the coefficients $6.0$ and $3.1$ are the number of events expected from the Southern sky in the case of normalization equal to one at 100 TeV, respectively for spectral index $\alpha=2$ and $\alpha=2.7$. The result is $F_{\mbox{\tiny G}}=2.5^{+2.4}_{-1.3}$ at 68$\%$ confidence level. 
Summarizing, with present information, 
the best model two-components flux is, 
\begin{equation} \label{chefig}
\frac{d \phi_{\mbox{\tiny S}}}{d E_\nu}=\bigg[ 2.5  \ \bigg(\frac{E_\nu}{100 \rm \ TeV} \bigg)^{-2.7} +2.1  \ \bigg(\frac{E_\nu}{100 \rm \ TeV} \bigg)^{-2} \bigg]  \cdot 10^{-18} \frac{1}{\rm GeV \ cm^2 \ sec \ sr}
\end{equation}
compared with the fluxes from Northern and Southern sky measured by IceCube, is shown in the Fig.~\ref{eeg}. \textcolor{black}{We show in the Fig.\ref{eeg} only the flux that describes the passing muons, that as discussed above, 
has a much smaller uncertainty than the flux that describes the HESE from Northern sky, and it is perfectly compatible with the latter.}
The red band is simply obtained propagating the uncertainties on the normalization, since the shape is fixed.
From this figure, the following two features are evident: 
At low energies, there is a good agreement between the two-components flux and the flux already measured from the Southern sky. At high energy, instead, there is a good agreement 
between the two-components flux and the flux already measured from the Northern sky.

Using the best fit values, the number of events from the Southern sky due to the extragalactic component is 12.6 (56.5$\%$), whereas the number of events due to the galactic component is 7.7 (34.5$\%$). \textcolor{black}{The expected number of events are calculated using the IceCube effective areas $A_{eff}^\ell$ (where $\ell$ denotes the flavor) of the Southern sky \cite{ice3}, and the fluxes found in this section, separating the contribution of the extragalactic and galactic component as follow:
\begin{equation}
N_{\mbox{\tiny EG,G}}=2 \pi \ \mbox{T}   \int_0^\infty \frac{1}{3} \frac{d\phi_{\mbox{\tiny EG,G}}}{dE_\nu} (A_{eff}^e+A_{eff}^\mu+A_{eff}^\tau) dE_\nu
\end{equation} 
where T=4 years is the exposure time.
We suppose that the flux for each flavor is 1/3 of the total flux, as expected in good approximation from a production mechanism due to  pion decay, after the occurrence of neutrino oscillations.}

Due to the great uncertainties on the extragalactic component, i.e. the flux from the Northern sky, the normalization of the galactic component (at 100 TeV) can fluctuate inside the interval between 1.2 - 4.9 giving a contribution between 17$\%$ and 67$\%$, to the number of events observed from the Southern sky. In table \ref{default} the number of events from the Southern sky observed by IceCube in 4 years is compared with theoretical prediction of the two-components neutrino flux. 

\subsubsection{Test with IceCube\label{angola}}

\textcolor{black}{In our hypothesis, a large fraction of the low energy events seen by IceCube have to come from a region close to the Galactic disk. 
The primary test is just the study of the angular distribution of the  HESE events detected by IceCube 
in the Southern sky, that is the region where galactic events are supposed to lie.}

\begin{figure}[t]
\centering
\includegraphics[scale=0.5]{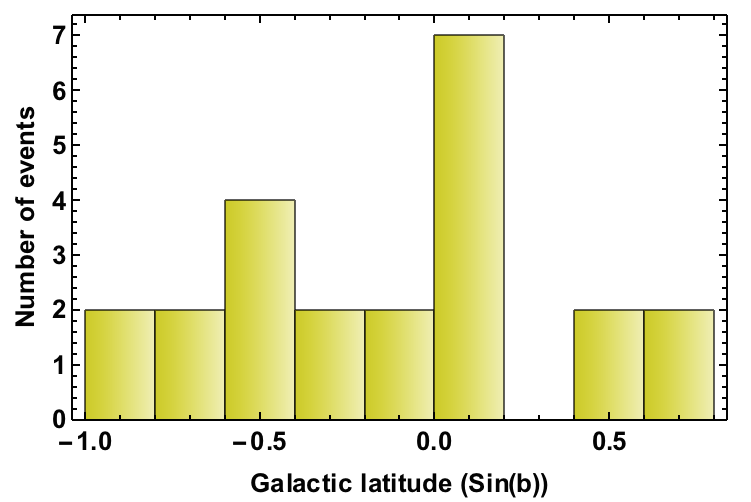}
\includegraphics[scale=0.52]{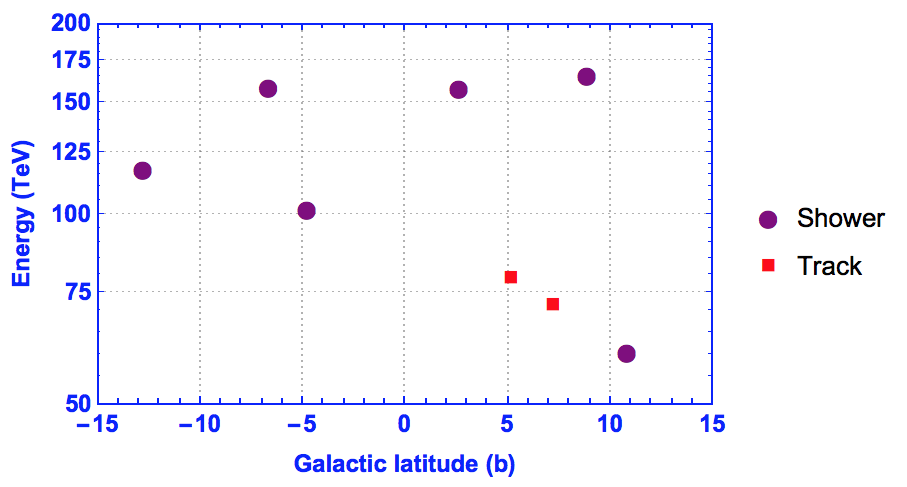}
\caption{\textit{Two presentations of events seen by IceCube in the Southern sky and with deposited energy above 60 TeV.
In the left panel we show the galactic latitude of all events.   
In the right panel we show those with energy below 200 TeV and in the region close to the Galactic plane.}}
\label{maf}
\end{figure}

\textcolor{black}{We begin by describing the model for the angular distribution.
We use the galactic coordinates $b$ (latitude) and $\ell$ (longitude) to identify the events, and consider the distribution in the  
variable 
\begin{equation}
x=\sin(b)
\end{equation}
First of all, we are interested in considering one population of galactic neutrino events 
characterized by a low  galactic latitude and localized mostly in the Southern sky. 
We assume the (normalized) differential distribution,
\begin{equation}
\lambda(x)= \frac{e^{-\frac{x^2}{2\delta x^2}}}{\sqrt{2\pi}\ \delta x}
\mbox{ with } \delta x=\sin 10^\circ
\end{equation}
where we have fixed the value of 
 angular extent of 
this population, 
$\delta x$,  
in view  of the angular resolution for shower events of IceCube.
For isotropic events instead, we expect a flat distribution in 
$x$ but localized in the Southern sky.
So we have, 
${dN}/{d\Omega}= \Theta( {\bf s}{\bf u})/(2 \pi)$, where 
 $d\Omega=d\ell\ d\sin(b)$, 
$\Theta$ is the Heaviside function, 
 ${\bf u}=(\cos b \cos \ell, \cos b \sin \ell, \sin b)$ is the direction of the event in galactic coordinates,
  ${\bf s}$ is the direction of the celestial South pole.\footnote{Its galactic latitude
   $b_{\mbox{\tiny CS}}$  equals the declination of galactic South (in  equatorial coordinates)
$b_{\mbox{\tiny CS}}=\delta_{\mbox{\tiny GS}}=-27.13^\circ$.}
  Integrating ${dN}/{d\Omega}$  over $\ell$ we get
\begin{equation}
\mu(x)=\frac{1}{\pi} \mbox{Re}\left[ \arccos\left(- \frac{x}{\sqrt{1-x^2}}\times \tan(b_{\mbox{\tiny CS}}) \right) \right]
\mbox{ with }
\tan(b_{\mbox{\tiny CS}})=-0.512
\end{equation}
We can use the normalized angular distribution 
\begin{equation}
\rho(x,f)=f \ \lambda(x)+(1-f)\ \mu(x) \mbox{ with }0\le f\le 1
\end{equation}
to model the cases 
where one or both components are present, depending on the value of  the fraction of events due to the galactic component,  $f$.}

\begin{figure}[t]
\centering
\includegraphics[scale=0.58]{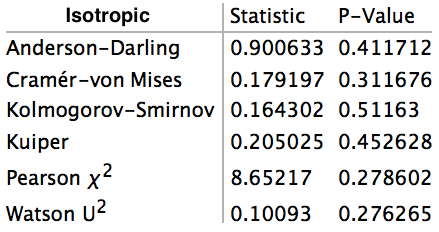}
\includegraphics[scale=0.58]{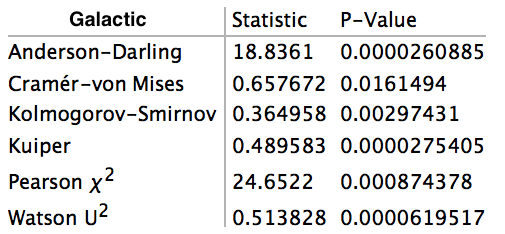}
\includegraphics[scale=0.58]{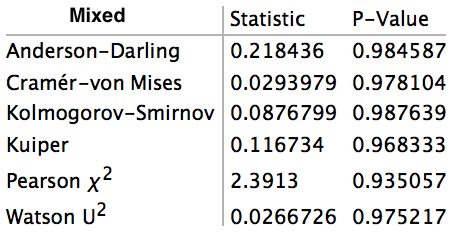} \\
\includegraphics[scale=0.55]{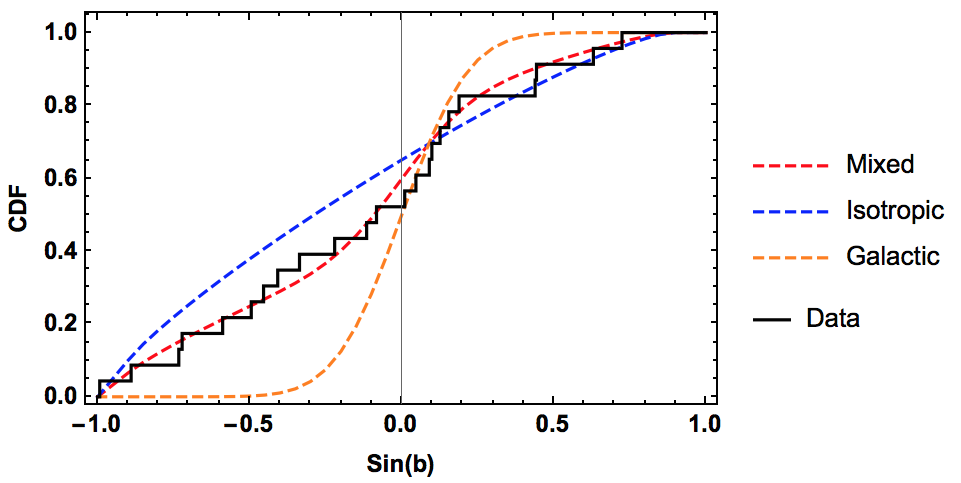}
\caption{\textit{Results of different hypothesis tests for isotropic, galactic and mixed model. In the bottom of the panel, the comparison between cumulative density function of the three models and data of Southern sky above 60 TeV. }}
\label{test}
\end{figure}

The data of IceCube (of which we present some graphical summaries in Fig.~\ref{maf}) can be compared with this 
model. The results are as follows,\\
a) When we fit to the data using the $\chi^2$ described in the Appendix, we find  $f=0.26 \pm 0.15$.
This is consistent with the previous independent determination obtained with the spectral information, that gave $f=0.35^{+ 0.32}_{-0.18}$. Note that with the angular analysis the uncertainty decreases.  \\
b) Moreover, we have tested the goodness of fit  for the three 
hypothesized angular distributions: 1) the extragalactic (isotropic) distribution
2) the purely galactic (low latitude distribution)
3) the mixed model (with both components).
It should be noted  that the model with mixed composition is not the result of a
fit to the angular data, but it is determined independently from 
the fit of the energy spectrum.
We find that all statistical tests in Fig.~\ref{test} give very similar results. A purely galactic component can be excluded. On the other side, \textit{both the isotropic and the mixed model are compatible with the data, although the model with two components fares better}. \textcolor{black}{The results of the hypothesis tests are reported in the table of Fig.~\ref{test}, where also a plot with the different cumulative distributions, compared with the experimental one, is shown.}\\
c) Finally, we note that our hypothesis implies a correlated angle-energy distribution in IceCube, namely a large fraction of the low energy events seen by IceCube in the Southern sky 
have to come from a region close to the Galactic disk.
The distribution of the data, given in
Fig.~\ref{maf}, shows that this is indeed the case. In fact, there are 8 events below 200 TeV {\em and} with $|b|<15^\circ.$ This is not so far to the expectation of 7.7 events, that one reads from Tab.~\ref{default}.

{\color{black}
Summarizing, we have performed various independent tests and analyses of  
the angular distribution of the events observed by IceCube. 
Our results are slightly more conservative but consistent with those obtained by  \cite{Troitsky:2015cnk} and do not contradict the ones of 
\cite{Neronov:2015osa}, 
see Sect.~\ref{boranga}.
The main findings are three,
1) The angular distribution of the Southern sky (HESE) events is 
in excellent agreement with the hypothesis of a  two-component distribution, formulated previously.
2) A purely extragalactic (isotropic) distribution cannot be excluded by the existing observations of the angular distribution.
3) The fraction of galactic neutrino events can be independently determined using only the spectra or using only the angular distribution: the two determinations are in agreement, and the second method gives a better estimation of the galactic contribution, because the uncertainty is smaller. 
The interest to increase significantly the present statistics is quite evident. }

\subsubsection{Test with ANTARES}
The data of ANTARES are important to probe the hypothesis of 
a Galactic component, as remarked in \cite{spurio}. 
A search  of an excess of through-going muons 
was performed using 1288 days of data, in the region   $|b|<4.5^\circ$ and  $|l|<39^\circ$ around the 
plane of the Milky Way \cite{visTesi,visProc}. 
A non-significant excess was seen and a bound 
was thus derived. Since this 
bound is 1.9 times tighter than the signal expected
{\em assuming that all IceCube events come from this region} (see Eq.~5.25 in  \cite{visTesi}) this extreme hypothesis is disfavored by ANTARES results. 

In our hypothesis only a fraction of the high energy neutrino signal derives from the region in the vicinity of the Galactic plane. An approximate Gaussian description of the results of \cite{visTesi} is given by $\chi^2(f)\approx 30 (f-1/6)^2$, where $f$ is the fraction of the events in the search region used by ANTARES. 
There is no contradiction with the hypothesis  that 
a large part of the Galactic neutrino signal 
is contained in the region investigated by ANTARES.
Moreover, the total fraction of Galactic events could be larger if the emitting region is larger. 
It is extremely interesting to continue 
the search for a diffuse neutrino signal 
in surroundings of the Galactic plane.

\subsection{Extragalactic component}
In this section, we show 
hypothesis of an extragalactic flux of neutrino $E^{-2}$ is in good agreement with the theory and with the observations.

\subsubsection{Comparison with the theory}
We are not interested to  consider a specific model here, but 
it is useful to make reference to a recent work in this field, where the plausibility of the the above hypothesis is clearly assessed. In the paper of Kalashev et al.~\cite{Kalashev:2014vya} is shown that a power law spectrum $E^{-2}$, in an energy range between $E_{\rm min} \simeq 100\ \rm TeV$ and some PeV, can be obtained in the context of AGN's. At the hearth of an AGN resides a super-massive black hole, surrounded by an hot accretion disc that emits thermal radiation. The AGN's can accelerate protons up to highest energies and photo-nuclear reactions with subsequent neutrino emission can occur. Considering the interaction between accelerated proton and a radiation with energy of $10^2-10^3$ eV, it is possible to obtain an $E^{-2}$ flux of neutrinos, as illustrated in the Fig.4 of \cite{Kalashev:2014vya}, where this kind of models are 
exhaustively discussed.

\begin{figure}[t]
\centering
\includegraphics[scale=0.55]{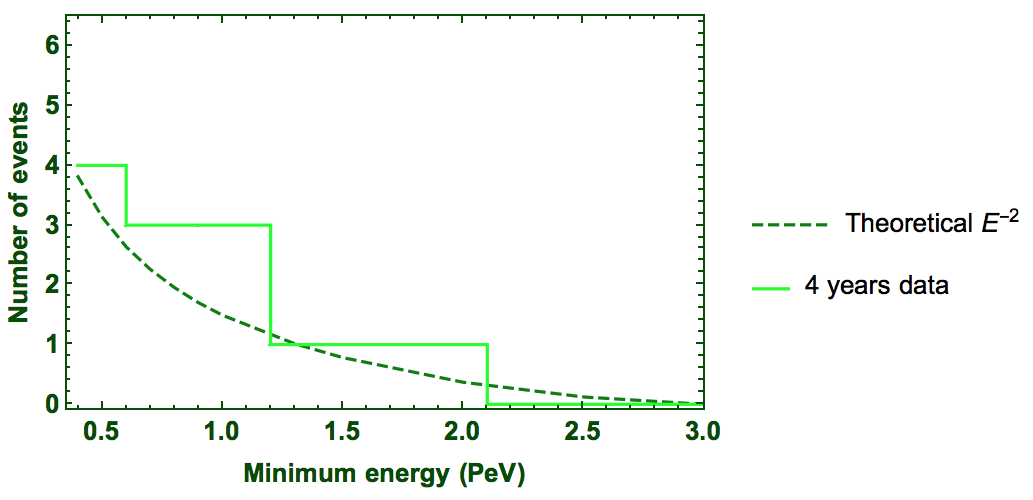}
\caption{\textit{Comparison between theoretical prediction and data observed between a minimum energy and 3 PeV. The line represents the energy in which the number of events between $E_{\rm min}$ and 3 PeV decreases by 1.}}
\label{datapred}
\end{figure}

\subsubsection{Comparison with the observations at the highest energies}
In order to check the validity of our assumption, the theoretical prediction has to be compared with the observed data at high 
energy, where the extragalactic component dominates. Let us analyze the number of showers, 
observed by IceCube in 4 years, between a minimum energy $E_{\rm min}$ and a certain maximum energy. The latter is chosen to be 
$E_{\rm min}=3$ PeV, in order to include all the events observed by IceCube and to avoid the Glashow resonance, that will be discussed in a separate section. 
In order to analyze events, it
is convenient to examine the cumulative function,
that is more appropriate to analyze sparse events. 
The comparison with the predictions can be done using the procedure described in \cite{Palladino:2015uoa}. Let us recall that the showers are mostly given by CC interactions of $\nu_e$ and $\nu_\tau$. The theoretical prediction are given by the formula:
\textcolor{black}{
\begin{equation}
N_S(E_{\rm min})=4 \pi \  \mbox{T} \int_0^{\rm 3 \ PeV} \frac{1}{3} 
\frac{{d \phi_{\mbox{\tiny EG}}}}{d E_\nu} A_{eff}^\tau \ [r(E_{\rm min}, E_{\nu})+r(E_{\rm min} \times 0.8, E_{\nu})] \ dE_{\nu}
\end{equation}
where $r(E_{\rm min}, E_\nu)$ is an energy resolution function defined in \cite{Palladino:2015uoa}  and each flavor carries a fraction
1/3 of the total flux  ${{d \phi_{\mbox{\tiny EG}}}}/{d E_\nu}$
defined previously. In this case the effective areas are the ones over 4$\pi$ and not the effective areas of the Southern sky.}
The results are illustrated in  
Fig.~\ref{datapred} where the minimum energy is 400 TeV, namely already in the region where the extragalactic component dominates. 

The spectrum $E^{-2}$ does not contradict the data at very high energy and fits well the experimental data until 3 PeV. A softer spectrum works well at low energies but it is in trouble to explain the 3 events above 1 PeV. 
Let us remark that with the cumulative analysis there is no lack of events between 0.5 PeV and 1 PeV, a feature that has been emphasized in many theoretical works. The lack is simply related the the choice of the bins and the low statistic of the events and for this reason it is not necessary to introduce an artificial component to explain a deficit of events. 

\section{Predictions \label{pre}}

We summarize our hypothesis. 
The high energy neutrinos are distributed according to  
a double component, with a softer spectrum below some 100 TeV, where the galactic component dominates, and an harder spectrum above $\simeq 0.5$ PeV, where the extragalactic component gives the main contribution. The latter is distributed as $E^{-2}$ while the second is distributed as $E^{-2.7}$. For what concerns the angular distribution, the  extragalactic component is distributed  isotropically whereas the galactic component is distributed around the Milky Way center and plane; the precise distribution (or, in first approximation, 
the extension of this component galactic latitude and longitude)
will be matter of observational investigations. 

There are many interesting consequences of this hypothesis,
for instance,\\
1) In the next years 
an event at Glashow resonance due to the extragalactic component
must be observed. \\
2) The observation of double pulse events will be the definitive proof of the cosmic origin of the high energy neutrinos. 
Evidently, this will be very important, even if not a specific test of our hypothesis.\\ 
3) The difference of the track-to-shower ratio of the two-components can be measured in IceCube.\\
4) There must be also high energy gammas associated to the galactic neutrinos. \\
5) In order to measure precisely the galactic component at low energy, an important contribution could come from telescopes located in the Northern Hemisphere, that has the possibility to measure the neutrinos from the Southern sky by means of passing muons, namely in a more clean way than the high energy starting events, that are instead  the observational tool exploited by IceCube. \\
We will discuss these tests of validation in this section.
\subsection{Showers above 3 PeV and Glashow resonance}
Using our estimation of the extragalactic flux, in 4 years 0.9 events due to \textcolor{black}{deep inelastic scattering} and 0.7 due to Glashow resonance ($p\gamma$ hypothesis) are expected. That means a total of 1.6 events, not yet observed by IceCube. At the present, this theoretical excess is not important and its significance is of 1.3$\sigma$. 
Following the approach of \cite{Palladino:2015uoa} it is possible to evaluate the number of years required to observe a shower above 3 PeV with a probability greater than 95$\%$, that corresponds at 2$\sigma$ in a Gaussian approach. In the case of $p\gamma$ mechanism of production, the number of events per year is about $\mu=0.4$. Using the Poissonian statistic, the probability to observe at least an event is given by $P(n>0)=1-\exp(-\mu)$, that means about 8 years for a probability greater than 95$\%$. So four more years are necessary to confirm our hypothesis or to disfavor it at 2$\sigma$ of significance.  It is important to notice that in the case of $\alpha=2.5$ more than 40 years are required to observe at least one shower above 3 PeV, with a $\mbox{C.L.}\geq95\%$; so there is an important difference between the single power law predictions and the two-components spectrum prediction. The contribution of the galactic component to this class of events is negligible, due to the harder spectrum.

\subsection{Double pulse}
The double pulse is generated by a $\nu_\tau$ that interact in CC, as clarified in \cite{Aartsen:2015dlt} and further quantified in  \cite{Palladino:2015uoa}. In the hypothesis of $p\gamma$ mechanism of production, neutrino oscillation give an about equal amount of $\nu_e$, $\nu_\mu$ and $\nu_\tau$ at the Earth. 
Our assumption on the extragalactic component of neutrinos implies that the expected number of events per years is $\simeq 0.12$. Using the same considerations of the previous section, about 20 years are required to observe a double pulse with a $\mbox{C.L.} \geq 95\%$ and with nowadays technology. Recall that the double pulse is less sensible to the shape of the neutrino spectrum {with respect to the Glashow resonance events and cascades above 3 PeV}, because about half of the signal is given by the already observed neutrinos, with energy between 100 TeV and 1 PeV \cite{Palladino:2015uoa}.  
{Note that the galactic component gives a small contribution to the double pulse events, i.e., $\simeq 0.03$ expected events per year. }

\subsection{The track-to-shower ratio}
\begin{figure}[t]
\centering
\includegraphics[scale=0.5]{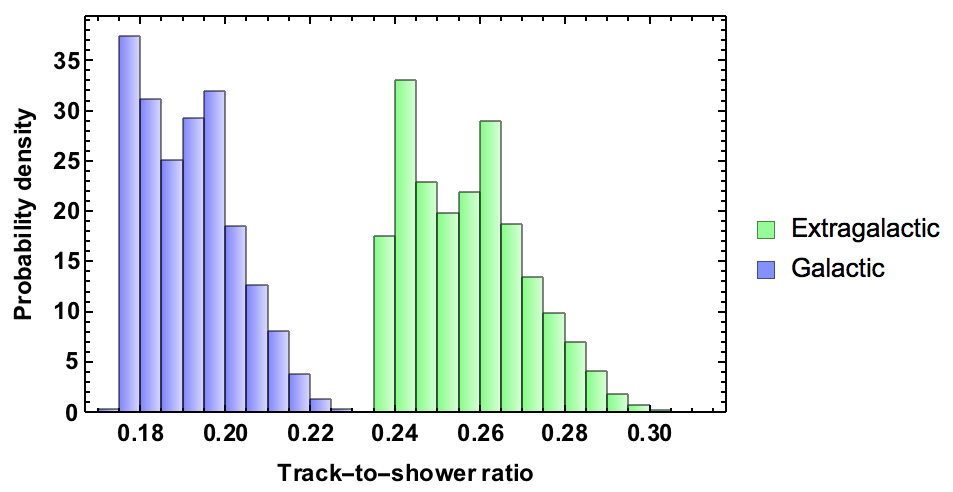}
\caption{\textit{Comparison between the track-to-shower ratio of galactic and extragalactic components, assuming the same mechanism of production, i.e., pion decay.}}
\label{kalAGN2}
\end{figure}

It is important to notice that the track-to-shower ratio for the galactic and the extragalactic components are different, even if the mechanism of production is the same -we refer to the standard case of pion decay in our paper-. This remains true despite the uncertainties on the oscillation parameters, as can be seen in the Fig.\ref{kalAGN2}. This difference is due to the different spectral shape, since the larger spectral index (namely the case of galactic $\nu$) penalizes track events.

\subsection{Implication for $\gamma$ astronomy}

The galactic component of neutrino implies the existence of a galactic component of very high energy gamma. These gamma rays can be observed, because their path in the Galaxy is less than their mean free path. The same consideration is not true for the extragalactic gamma rays associated to neutrinos, because when the energy is  above 10 TeV they are absorbed after about 100 Mpc, due to the interaction with the extragalactic background light (EBL). It is possible to give a lower limit on the flux of gammas, that corresponds to the case of $pp$ interaction at the source. 
The ratio between the flux of gamma and the flux of neutrinos is given by the following equation based on \cite{Villante:2008qg},
\begin{equation}
\frac{\phi_\gamma}{\phi_\nu}=1.009 + 0.622 (\alpha-2.3) +0.140 (\alpha-2.3)^2 +0.009 (\alpha-2.3)^3 
\end{equation}
that takes the value 
${\phi_\gamma}/{\phi_\nu}= 1.28$ if $\alpha=2.7$.
Therefore, the flux of galactic gamma must be of the same order of the neutrino flux, more precisely:
\begin{equation}
\phi_\gamma (E) \geq 1.28 \ \phi_{\mbox{\tiny G}}(E)
\end{equation}
Let us notice that the cutoff on the $\phi_\gamma$ is at higher energy with respect to the cutoff of the galactic neutrino spectrum. In fact the energy of $\gamma$ is typically $E_\pi/2$, whereas the energy of neutrinos is $\simeq E_\pi/4$ (see \cite{Kelner:2006tc} for a complete description of the secondary neutrinos spectrum). The relationship between galactic HE neutrinos and HE $\gamma$-rays is also discussed in \cite{Sahakyan:2015bgg}, focusing on SNRs, PWN and binary systems. Recall also that in the case of binary system if the $\gamma$-rays are produced from protons, the flux of HE neutrinos can significantly exceed the one of $\gamma$-rays (by a factor of $e^\tau$ where $\tau$ is the optical depth) considering that they escape from the region without absorption, unlike the $\gamma$-rays. 

\begin{figure}[t]
\centering
\includegraphics[scale=0.6]{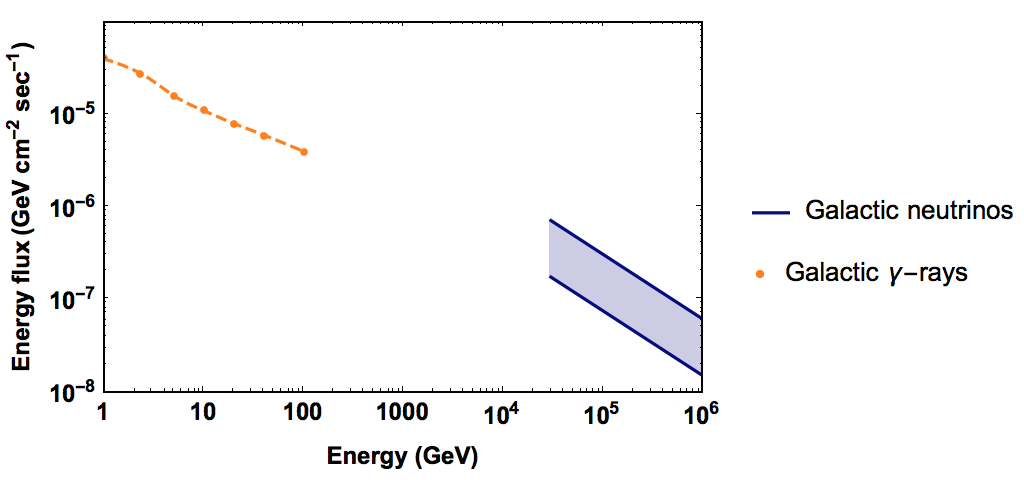}
\caption{{\textit {In orange, spectrum of diffuse gamma-ray emission from the inner Galaxy ($-80^o \leq l \leq 80^o$, $-8^o \leq b \leq 8^o$, measured by Fermi-LAT. In blue, spectrum of galactic neutrinos with their uncertainties.}}}
\label{gammagal}
\end{figure}

In the other case,  the flux of gamma must be greater than the flux of neutrinos to have a consistent theory. At the present there are no measurements of  the gamma rays in the same energy range of neutrinos (above 30 TeV), so a direct comparison is not possible. Fig.~\ref{gammagal} shows the diffuse flux of galactic gamma rays (see \cite{Abdo:2009mr},\cite{Huntemeyer:2010zz}) and the estimated flux of galactic neutrinos in the energy range in which they are already measured. No firm conclusion can be derived nowadays 
from these results. 

\subsection{Future km$^3$-class telescopes in Northern Hemisphere}


Our hypothesis requires that in the vicinity of the galactic disk, there is a population of high energy neutrinos with energies below few 100 of TeV. The present data are compatible with a large fractions of events from this origin. In our reference hypothesis, the fraction is  about  30$\%$ but with a great uncertainties, since the 1$\sigma$ interval includes values that differ by a factor of two (above or below). 
This means that the existence of a galactic component is 
reasonable even if nowadays it is hard to quantify it precisely; future experiments in the Northern hemisphere will be necessary to prove its existence and to measure this component precisely.

Let us remark that the $\simeq 30 \%$ is referred to the fraction of supposed galactic events with respect to the total number of events observed from the Southern sky. If all the events (both from South and North) are considered, 
the galactic contribution becomes 24$\%$ at best fit. However let us repeat that the precise extension of this signal (in angle and in energy) is not known and this requires further experimental investigations. 
In order to validate this assumption, neutrino telescopes of sufficient volume and located in the Northern hemisphere are needed. 

The KM3NeT \cite{km3}
and the BAIKAL-GVD \cite{gvd}
are such kind of experiments; the first will be in the Mediterranean Sea, the second one in Baikal Lake.
They will be in the right geographical position to validate the presence of a diffuse galactic neutrino component, thanks to the possibility to use passing muons of relatively low energies. Note that in water the angular resolution is much better than in ice and it attains sub-degree \cite{km3}. In this manner, the precise angular extension will be investigated.  Moreover, also `high energy starting events' 
below few 100 TeV will be useful for the same purpose.

Also the extragalactic component can be measured with a similar precision of the IceCube experiment, using the high energy starting events in the energy range above some 100 TeV, where the background due to atmospheric neutrinos is negligible.

\section{Discussion and summary}
\label{summary}
The main goal of IceCube has been to show conclusively that high-energy neutrinos exist. To consolidate this goal, a simple assumption proved to be reasonably appropriate to summarize the data and the present understanding, namely, the assumption of a single new population, distributed isotropically and with a power law spectrum with slope significantly different from $\alpha=2$.
This is a minimal modification of the hypothesis suggested by the scenario of extragalactic origin but it leads to a certain number of issues, examined and discussed in Sect.~\ref{cd}.

However, there is no strong theoretical reason to omit the galactic contribution to the observed high-energy neutrinos; on the contrary, astronomical data, gamma rays included, suggest including it. We have shown that the inclusion of another significant component of  the high energy neutrino flux, with the slope of the observed galactic cosmic rays spectrum, and with an extension till few 100 TeV, is consistent with the known IceCube data.  

The angular distribution of the total galactic contribution (including point sources and diffuse component) is not precisely known but the galactic latitude should be within $|b|<(5-20)^\circ$ and the galactic longitude extends from $|l|<40^\circ$ or an even wider region.
Although we would like to observe point sources eventually, the high energy neutrino astronomy is in a similar position as the ordinary one more than 200 years ago, at the times of T.~Wright, I.~Kant, W.~Herschel, when the shape of the Milky Way was understood. It should be stressed, however, that neutrinos will give unique information on galactic and extragalactic cosmic rays. 

We have shown that the dominant extragalactic component predicts observable events of new type in IceCube, those from Glashow resonance and those of  double-pulse type. We have emphasized that the galactic component 
is more important at comparably low energies (below few 100 TeV)  and can be observed by exploiting through-going muons from the Northern Hemisphere.  It leads to a peculiar flavor ratio and should be accompanied by a (diffuse) emission of very high energy gamma rays, calculable from the neutrinos, and extended till some few 100 TeV.

In the near future, a combined analysis of the data of IceCube and ANTARES  will permit us to obtain more precise inferences on the parameters of the two component model. It is important to emphasize that within this model, the angular features and those of the  energy spectrum are closely entangled and should be analyzed together.


\subsection*{Acknowledgments}
We would like to thank
F.~Aharonian,
R.~Aloisio,
P.~Lipari 
for pleasant and useful discussions and 
R.~Coniglione, 
A.~Esmaili,
G.~Pagliaroli,
P.~Sapienza, 
F.~L.~Villante 
for collaboration on related subjects.

\appendix

\textcolor{black}{\section{Statistical procedures}
In the main text, we considered various questions:
1) We want to test whether two distributions, assumed to be power law, but measured by means of two different datasets, are compatible among them or not.
2) We want to study the angular distribution of the events from the Southern sky and to obtain  the optimal combination of galactic and extragalactic neutrino fluxes from the angular distribution of the data.
Below, we describe the procedures we have followed to answer them.}

\textcolor{black}{\subsection{Compatibility of two measurements}
We have two measurements of one quantity $x$ 
that are summarized by two likelihood functions $\ell_1(x)$ and $\ell_2(x)$. We want to test whether the two measurements  
are compatible or not. So, we introduce the variable  that quantifies the 
difference of values $\delta x$ and associate to this variable 
a new likelihood function by mean of the following overalap integral,
\begin{equation}
\ell\left(\hskip0.5mm \delta x \hskip0.5mm\right)=\int_{-\infty}^{+\infty} dx\ \ell_1(x)\, \ell_2(x-\delta x)
\end{equation}
where the integrals are taken over the range of variation of the variable $x$ that we assume to be the whole real axis.
In the common case when the two likelihoods are approximated by Gaussian functions, 
$\ell_i(x)=g(x,\mu_i,\sigma_i)$, where,
\begin{equation}
g(x,\mu,\sigma)=\frac{e^{-\frac{(x-\mu)^2}{2 \sigma^2}}}{\sqrt{2\pi} \sigma} 
\end{equation}
it is straightforward to show that,
\begin{equation}
\ell\left(\hskip0.5mm \delta x \hskip0.5mm\right)=g\left(\delta x\ ,\ \mu_1-\mu_2\ ,\ \sqrt{\sigma_1^2+\sigma_2^2}\right)
\end{equation}
This result is very reasonable and admits the following interpretation: 
The value $\delta x=0$, that can be considered as the case when there is no difference between the two likelihoods, deviates from the best fit value by the following `numbers of sigma',
\begin{equation}
N_{\sigma}=\frac{|\mu_1-\mu_2|}{\sqrt{\sigma_1^2+\sigma_2^2}}
\end{equation}
Supposing $\mu_1> \mu_2$ 
(and quite similarly if $\mu_1< \mu_2$) 
we  can use  the tail of the likelihood $\ell$ to estimate the 
statistical significance of a similar (or stronger) discrepancy,
\begin{equation}
\mbox{p-value}=2 \int_{-\infty}^0 dy\ \ell(y)
\end{equation}
that corresponds to $N_\sigma$ if the Gaussian case applies and 
of course should be used only when the outcome is less than 1.
This quantifies the significance of null hypothesis, that the two  measurements are compatible. If the function $\ell$ does not deviate strongly from a Gaussian distribution, we can generalize the procedure and use the above p-value to quantify the difference 
between two likelihood functions with different  medians $\mu_1>\mu_2$. }

\textcolor{black}{\subsection{Analysis of a two component model}
Suppose that we have a set of measurements of a  
certain observable quantity $x$, with values $x=x_1,x_2,x_3...$, 
and to know that they come from two different populations, with known distributions  
$\lambda(x)$ and $\mu(x)$, normalized to 1. We ask which is the 
optimal combination of  these distributions 
that reproduces the set of measurements. This can be answered introducing the 
fraction of events of the first type $0\le f\le 1$, corresponding to a fraction of $1-f$ events of the second type, 
and using the following $\chi^2$,
\begin{equation}
\exp\left(-\frac{\chi^2(f)}{2} \right)=\prod_{i}\ [\, f \lambda_i + (1-f) \mu_i \, ] 
\mbox{ where }\lambda_i=\lambda(x_i)\mbox{ and }
\mu_i=\mu(x_i)
\end{equation}
This $\chi^2$ can be used to estimate the 
fraction $f$.
Suppose there is a minimum for some value $f=\bar{f}$ internal to the physical interval 
$0<\bar{f}<1$ and such that 
\begin{equation}
\chi^2(f)=\mbox{constant}+
\frac{(f-\bar{f})^2}{\delta f^2}+ o((f-\bar{f})^2)
\end{equation}
 it is easy to obtain the following analytical expressions,
\begin{equation}
\sum_{i} \xi_i(\bar f)= 0 \ \mbox{ and }\ 
\delta f^2=\frac{1}{\sum_{i} \xi_i(\bar f)^2}  
\mbox{ where } \xi_i(f)= \frac{\lambda_i-\mu_i}{\mu_i +  f (\lambda_i-\mu_i)} 
\end{equation}
The two quantities $\bar{f}$ and $\delta f$ characterize which is the optimal combination 
of the two populations that resembles more closely  the given set of measurements, 
and they are usually quoted as $f=\bar{f}\pm \delta f$.}

\end{document}